\documentclass[a4paper,twocolumn,11pt,accepted=2021-06-15]{quantumarticle}
\pdfoutput=1 
\usepackage[utf8]{inputenc}
\usepackage[english]{babel}
\usepackage[T1]{fontenc}
\usepackage{amsmath}
\usepackage{amssymb}
\usepackage{cite}
\usepackage{bbm}
\usepackage{color}
\usepackage{xcolor,hyperref}
\usepackage{graphicx}
\usepackage{tikz}
\usepackage{lipsum}
\newcommand{\naw}[1]{\left(#1\right)}
\newcommand{\ket}[1]{\left|#1\right>}
\newcommand{\bra}[1]{\left<#1\right|}
\newcommand{\av}[1]{\left<#1\right>}
\newcommand{\com}[1]{\left[#1\right]}
\newcommand{\modu}[1]{\left|#1\right|}
\newcommand{\poisson}[1]{\left\{#1\right\}}

\begin{document}

\title{Classical and quantum speed limits}
\author{Katarzyna Bolonek-Laso\'n}
\affiliation{Department of Statistical Methods, Faculty of Economics and Sociology University of Lodz, 41/43 Rewolucji 1905 St., 90-214 Lodz,  Poland}
\author{Joanna Gonera}
\author{Piotr Kosi\'nski}
\thanks{piotr.kosinski@uni.lodz.pl}
\affiliation{Department of Computer Science, Faculty of Physics and Applied Informatics University of Lodz, 149/153 Pomorska St., 90-236 Lodz, Poland}

\maketitle
\begin{abstract}
The new bound on quantum speed limit (in terms of relative purity) is derived by applying the original Mandelstam-Tamm one to the evolution in the space of Hilbert-Schmidt operators acting in the initial space of states. It is shown that it provides the quantum counterpart of the classical speed limit derived in \textit{Phys. Rev. Lett. \textbf{120} (2018), 070402} and the $\hbar\rightarrow 0$ limit of the former yields the latter. The existence of classical limit is related to the degree of mixing of the quantum state. 
\end{abstract}

\section{Introduction}\label{I}
There exist two seminal results concerning the bounds on the speed of quantum evolution and related ability to distinguish quantum states connected via time evolution. The first one, due to Mandelstam and Tamm, is expressed in terms of energy dispersion of initial state \cite{Mandel}. Quite unexpectably, Margolus and Levitin \cite{Margolus} established an independent bound based on the expectation value of excitation energy. Unifying both results one obtains the following constraint on orthogonalization time \cite{Levitin}
\begin{equation}
t_\perp\geq \max\naw{\frac{\pi\hbar}{2(\av{E}-E_0)},\frac{\pi\hbar}{2\Delta E}}\label{a1}
\end{equation}
with $E_0$ being the ground state energy while $\Delta E$ is the energy dispersion. This result has been further analyzed, extended in various directions and applied in different context in numerous papers \cite{Fleming, Wootters, Bhatta, Anandan, Vaidman,Uhlmann, Pfeifer, Lloyd, Giovannetti,Kosinski, Zielinski,Caneva, Mostaf,Jones, Campo,Demkowicz,Taddei, Campo1,Deffner,Poggi,Uzdin,Xu1,Zhang,Sun,Wu1, Marvian,Mondal,Uzdin1,Marivan1,Pires,Campbell, Campo2,Deffner1,Defner2,Cianciaruso, Cai,Campaioli,Ito, Campaioli1, Teittinen,Sun1, Diaz,Fogarty, Connor, Teittinen1}. 

Quite recently an interesting question has been raised \cite{Shanahan,Okuyama} whether there exists a classical counterpart of quantum speed limit. It has been further analyzed from various points of view in a number of subsequent papers \cite{Shiraishi, Funo, Vo, Wu2, Nicholson}.

Consider the $\hbar\rightarrow 0$ limit of the bound \eqref{a1}. Typically, both the excitation energy and its dispersion behave as $O(1)$ with $\hbar\rightarrow 0$. In fact, denoting generically the "principal" quantum number by $n$ one finds that energy level has an expansion of the form $a_0(n\hbar)+a_1(n\hbar)\hbar+a_2(n\hbar)\hbar^2+\ldots$. By correspondence principle the classical limit is obtained letting $\hbar\rightarrow 0$, $n\rightarrow \infty$, $n\hbar=O(1)$. Therefore, the right hand side of eq.~\eqref{a1} vanishes in this limit. This is quite reasonable. Consider the quantum pure states which saturate the Heisenberg uncertainty principle. The $\hbar\rightarrow 0$ limit yields the pure classical state $\rho_0$ with delta-like probability distribution. Clearly, the overlap between $\rho$ and $\rho_t$ vanishes for any $t\neq 0$ (except some "static" states). However, the question becomes more subtle if mixed states are taken into account.

In what follows we consider the Mandelstam-Tamm bound (the Margolus-Levitin one seems to be less interesting in the classical limit \cite{Okuyama}). It follows from the following inequality 
\begin{equation}
\modu{\av{\Psi(0)|\Psi(t)}}\geq\cos\naw{\frac{(\Delta E)_0}{\hbar}t}.\label{a2}
\end{equation}
Both eqs.~\eqref{a1} and \eqref{a2} show that the trouble with $\hbar\rightarrow 0$ limit results from the fact that $\frac{\Delta E}{\hbar}\rightarrow\infty$. So the question is whether the bound \eqref{a2} can be replaced by the one not involving the troublesome expression $\frac{\Delta E}{\hbar}$. If $\hat{\rho}=\ket{\Psi}\bra{\Psi}$ is a pure state one can immediately derive the identity
\begin{equation}
\frac{2}{\hbar^2}\naw{\Delta E}^2_\rho=-\frac{1}{\hbar^2}Tr[\hat{H},\hat{\rho}]^2.\label{a3}
\end{equation}
The right hand side involves the commutator. Therefore, one can expect that it possesses well-defined classical limit for "reasonable" class of density operators $\hat{\rho}$. This class obviously does not encompass the set of pure states since the left-hand side is not well-defined in the limit $\hbar\rightarrow 0$. However, the above identity is valid \textbf{only} for pure states; the more mixed is the state $\hat{\rho}$ the more the right-hand side of \eqref{a3} deviate from the left one.
Therefore, one can expect that for $\hat{\rho}$ describing sufficiently "regular" mixed states the right hand side behaves reasonably for $\hbar\rightarrow0$.
Consequently, the idea to modify the Mandelstam-Tamm bound to get sensible $\hbar\rightarrow 0$ limit by replacing somehow $(\delta E)^2_\rho$ by $Tr[\hat{H},\hat{\rho}]^2$ is appealing.

On the classical level the bound on the speed of Hamiltonian evolution in phase space has been derived by Okuyama and Ohzeki \cite{Okuyama}. They considered the Hilbert space of square integrable functions on phase space. By applying some Hilbert space techniques they derived a bound valid for square integrable classical probability distributions. They suggested that this bound is specific for classical dynamics since the generator of dynamics in phase space (Liouvillian) is a first order differential operator, contrary to the Schr\"odinger one. The bound derived by Okuyama and Ohzeki involves the Poisson bracket of Hamiltonian and probability density distribution. This suggests that it is the classical limit of some quantum bound involving the relevant commutator instead of $(\Delta E)_\rho$.

In the present paper we show that the relevant quantum bound can be readily obtained from \textbf{the Mandelstam-Tamm relation applied to the pure states in the Hilbert space of Hilbert-Schmidt operators acting in the original space of states}. 
Let us explain our idea more precisely. Given a quantum system we start with the Hilbert space $\mathcal{H}$ providing the description of its set of states. In general, the latter are represented by density matrices, i.e. positive semidefinite trace one operators acting in $\mathcal{H}$; in particular, the pure states correspond to onedimensional projectors. Now, the density matrices are Hilbert-Schmidt operators. It is well known that the set of H-S operators with the natural linear space structure and the scalar product defined by an appropriate trace (cf. eq.~\eqref{a20} below) form the Hilbert space $\mathcal{H}_{HS}$. Therefore, the states of the physical system under consideration, which are described by density matrices acting in the initial space of states $\mathcal{H}$, correspond to the vectors in $\mathcal{H}_{HS}$ (obviously, not all vectors in $\mathcal{H}_{HS}$ describe the states of our system). The dynamical evolution of density matrix $\hat{\rho}$ (given by quantum Liouville equation) can be equivalently described as unitary evolution of the vector $\hat{\rho}$ in $\mathcal{H}_{HS}$. Now, the Mandelstam-Tamm relation is the mathematical statement concerning one-parameter groups of unitary transformations in an arbitrary Hilbert space. Consequently, it can be applied to the evolution of the vector $\hat{\rho}\in\mathcal{H}_{HS}$. However, from the point of view of the initial space $\mathcal{H}$ $\hat{\rho}$ is a density matrix describing the actual, in general mixed, state of a physical system. As a result, istead of the usual inner product for state vectors in $\mathcal{H}$ the distance is defined by Hilbert-Schmidt inner product for density matrices.

The new bound is expressed in terms of relative purity and is tighter than some encountered in the literature (in the Appendix we compare the bound on quantum speed derived here with the another one used in \cite{Shanahan} in the context of classical limit).  Moreover, it is the quantum counterpart of Okuyama-Ohzeki classical bound. As it has been mentioned above their bound is valid for square integrable probability distributions. We show that this assumption implies that it is obtained from density operators describing the states which become more and more mixed as $\hbar\rightarrow 0$.

The paper is organized as follows. In Sec.~\ref{II} we rederive the Okuyama-Ohzeki bound showing that it is a direct consequence of Mandelstam-Tamm one given by eq.~\eqref{a2}. Then in Sec.~\ref{III} we derive the quantum bound by using \eqref{a2} in Hilbert space of Hilbert-Schmidt operators. We indicate also why one cannot expect the bound to be saturated. Sec.~\ref{IV} is devoted to the Wigner function formalism applied to speed limit. It is shown that the classical limit of the bound derived here yields Okuyama-Ohzeki bound. Sec.~\ref{V} contains short summary.  Finally, in the Appendix A we compare our bound with the one considered by Shanahan et al., also in the context of (semi)classical limit. Appendix B is devoted to some technicalities.   

\section{Speed limit in classical phase space}\label{II}
Let us rederive the bound on classical speed limit (CSL) obtained in Ref.~\cite{Okuyama}. Consider a classical dynamical system described by some $2f$ dimensional phase space $\Gamma$ and a Hamiltonian $H(\underline{q},\underline{p})$; for simplicity we assume that $H$ is time independent but the generalization is straightforward. Let $\rho(q,p,t)$ be a probability density of classical states. The classical Hamiltonian dynamics is encoded in Liouville equation
\begin{equation}
\frac{\partial\rho(\underline{q},\underline{p},t)}{\partial t}+\poisson{\rho(\underline{q},\underline{p},t),H(\underline{q},\underline{p})}=0.\label{a4}
\end{equation}
Eq.~\eqref{a4} is simply the conservation law for $\rho(\underline{q},\underline{p},t)$ along trajectories. Therefore, any differentiable function $G(\rho)$ obeys eq.~\eqref{a4} as well.

Eq.~\eqref{a4} is the classical counterpart of von Neumann equation with Poisson bracket replacing the commutator $\frac{-i}{\hbar}\com{\hat{\rho},\hat{H}} $. However, we would like to view it as defining one-parameter group of unitary transformations in an appropriate Hilbert space. To this end we rewrite it in the form 
\begin{equation}
\frac{i\partial\rho(\underline{q},\underline{p},t)}{\partial t}=(\hat{L}\rho)(\underline{q},\underline{p},t)\label{a5}
\end{equation}
where the Liouvillian $\hat{L}$ is defined by 
\begin{equation}
\begin{split}
(\hat{L}\rho)&(\underline{q},\underline{p},t)\equiv i\poisson{H,\rho}(\underline{q},\underline{p},t)=\\
& =i\sum_{k=1}^f\naw{\frac{\partial H}{\partial q_k}\frac{\partial\rho}{\partial p_k}-\frac{\partial H}{\partial p_k}\frac{\partial\rho}{\partial q_k}}(\underline{q},\underline{p},t)\end{split}\label{a6}
\end{equation}
i.e,
\begin{equation}
\hat{L}\equiv i\sum_{k=1}^f\naw{\frac{\partial H}{\partial q_k}\frac{\partial}{\partial p_k}-\frac{\partial H}{\partial p_k}\frac{\partial}{\partial q_k}}\label{a7}
\end{equation}
is a first order differential operator.

Eq.~\eqref{a5} has the form of Schr\"oedinger equation with $\hat{L}$ playing the role of Hamiltonian. Therefore, we consider the Hilbert space of functions $f(\underline{q},\underline{p})$, square integrable over the phase space $\Gamma$; the relevant scalar product is defined by
\begin{equation}
(f,g)\equiv \int\limits_\Gamma \mathrm{d}\underline{q}\,\mathrm{d}\underline{p}\,\overline{f(\underline{q},\underline{p})}\,g(\underline{q},\underline{p}).\label{a8}
\end{equation} 
It is easy to see that $\hat{L}$ is (at least formally) selfadjoint with respect to the above scalar product. Assuming that our probability distribution $\rho(\underline{q},\underline{p},0)$ is square integrable we can consider eq.~\eqref{a5} as describing an unitary evolution in our Hilbert space. Consequently the original Mandelstam-Tamm derivation remains valid.

 Let $\Vert\rho_t\Vert$ denotes the norm in our Hilbert space,
 \begin{equation}
 \Vert\rho_t\Vert^2\equiv \int\limits_{\Gamma}\mathrm{d}\underline{q}\,\mathrm{d}\underline{p}\,\rho^2(\underline{q},\underline{p},t);\label{a9}
 \end{equation}
 then $\frac{1}{\Vert\rho\Vert}\rho(\underline{q},\underline{p},t)$ is the normalized vector. The Mandelstam-Tamm bound, applied in this context, yields (cf.~\cite{Bhatta} for convenient form of MT bound)
 \begin{equation}
 \frac{(\rho_0,\rho_t)}{\Vert\rho_0\Vert^2}\geq\cos\naw{(\Delta L)_0t}.\label{a10}
 \end{equation}
 Note that due to
 \begin{equation}
 (\rho_0,\rho_t)\equiv \int\limits_{\Gamma}\mathrm{d}\underline{q}\,\mathrm{d}\underline{p}\,\rho(\underline{q},\underline{p},0)\,\rho(\underline{q},\underline{p},t)\label{a11}
 \end{equation}
$\frac{1}{\Vert\rho_0\Vert^2}(\rho_0,\rho_t)$ can be viewed as classical counterpart of relative purity \cite{Audenaert}, \cite{Frey}. The dispersion $(\Delta L)_0$ is defined as usual,
\begin{equation}
(\Delta L)_0^2\equiv\frac{1}{\Vert\rho_0\Vert^2}\naw{(\rho_0,\hat{L}^2\rho_0)-(\rho_0,\hat{L}\rho_0)^2}.\label{a12}
\end{equation} 
Now, let $K$ be the antiunitary operator of complex conjugation. Then, from eq.~\eqref{a6} one finds
\begin{equation}
K\hat{L}K=-\hat{L}.\label{a13}
\end{equation}
Moreover, $\rho$ is real, $K\rho=\rho$, and
\begin{align}
\begin{split}
(\rho,\hat{L}\rho)&=(K\rho,\hat{L}K\rho)=-(K\rho,K\hat{L}\rho)=\\
&=-(\hat{L}\rho,\rho)=-(\rho,\hat{L}\rho)\end{split}\label{a14}
\end{align}
and eq.~\eqref{a12} simplifies to
\begin{equation}
(\Delta L)_0^2=\av{\hat{L}^2}_0\equiv\frac{(\rho_0,\hat{L}^2\rho_0)}{\Vert\rho_0\Vert^2}.\label{a15}
\end{equation}
So, eq.~\eqref{a10} can be rewritten as
\begin{equation}
\frac{(\rho_0,\rho_t)}{(\rho_0,\rho_0)}\geq\cos\naw{\frac{\sqrt{(\rho_0,\hat{L}^2\rho_0)}}{\Vert\rho_0\Vert}\,t}.\label{a16}
\end{equation}
As it has been noted above, any function $G(\rho)$ of $\rho$ obeys Liouville equation and one can repeat the above reasoning. In particular, taking $G(\rho)=\rho^\alpha$ and assuming that $\rho^\alpha$ is square integrable one obtains
\begin{equation}
\frac{(\rho_0^\alpha,\rho_t^\alpha)}{(\rho_0^\alpha,\rho_0^\alpha)}\geq \cos\naw{\sqrt{\frac{(\rho_0^\alpha,\hat{L}^2\rho_0^\alpha)}{(\rho_0^\alpha,\rho_0^\alpha)}}\,t}\label{a17}
\end{equation}
which coincides with eq.~(18) of Ref.~\cite{Okuyama}. Note further that, by virtue of eq.~\eqref{a6},
\begin{equation}
(\rho_0,\hat{L}^2\rho_0)=(\hat{L}\rho_0,\hat{L}\rho_0)=\int\limits_\Gamma\mathrm{d}\underline{q}\,\mathrm{d}\underline{p}\poisson{H,\rho_0}^2.\label{a18}
\end{equation}

\section{Quantum speed limit}\label{III}
We shall now find the quantum counterpart of the bounds derived above. Let $\mathcal{H}$ be the Hilbert space of states of the quantum system under consideration. Any physical state is described by a density matrix $\hat{\rho}(t)$ obeying
\begin{equation}
i\hbar\dot{\hat{\rho}}(t)=\com{\hat{H},\hat{\rho}(t)}\label{a19}
\end{equation}
with $H$ being the Hamiltonian of the system under consideration.

Consider the Hilbert space $\mathcal{H}_{HS}$ of Hilbert-Schmidt operators acting in $\mathcal{H}$, equipped with the scalar product
\begin{equation}
(A,B)\equiv Tr(A^+B).\label{a20}
\end{equation}
Given the Hamiltonian $\hat{H}$ we define the operator $\tilde{H}$, acting in $\mathcal{H}_{HS}$, by
\begin{equation}
\tilde{H}A\equiv\com{\hat{H},A}\label{a21}
\end{equation}
$\tilde{H}$ is selfajoint with respect to the scalar product \eqref{a20}. Eq.~\eqref{a19} may be rewritten as follows
\begin{equation}
i\hbar\dot{\hat{\rho}}_t=\tilde{H}\hat{\rho}_t.\label{a22}
\end{equation}
Taking into account that $\frac{1}{\Vert\hat{\rho}_t\Vert}\hat{\rho}_t$ ($\Vert\rho_t\Vert\equiv\sqrt{Tr(\hat{\rho}^2_0)}$) is a unit vector in $\mathcal{H}_{HS}$ one can again apply Mandelstam-Tamm inequality to find 
\begin{equation}
\frac{Tr(\hat{\rho}_0\hat{\rho}_t)}{Tr(\hat{\rho}_0^2)}\geq\cos\naw{\frac{(\Delta\tilde{H})_0}{\hbar}\,t}\label{a23}
\end{equation}
with
\begin{equation}
(\Delta\tilde{H})^2_0=\frac{1}{\Vert\hat{\rho}_t\Vert^2}\naw{(\hat{\rho}_t,\tilde{H}^2\hat{\rho}_t)-(\hat{\rho}_t,\tilde{H}\hat{\rho}_t)^2}.\label{a24}
\end{equation}
Following the reasoning similar to that in previous section we define an antiunitary operator $K$, acting in $\mathcal{H}_{HS}$:
\begin{equation}
KA=A^+.\label{a25}
\end{equation}
Then
\begin{equation}
K\tilde{H}K=-\tilde{H}\label{a26}
\end{equation}
and
\begin{equation}
K\hat{\rho}_t=\hat{\rho}_t.\label{a27}
\end{equation}
Consequently
\begin{align}
\begin{split}
(\hat{\rho}_0,\tilde{H}\hat{\rho}_0)&=(K\hat{\rho}_0,\tilde{H}K\hat{\rho}_0)=-(K\hat{\rho}_0,K\tilde{H}\hat{\rho}_0)=\\
&=-(\tilde{H}\hat{\rho}_0,\hat{\rho}_0)=-(\hat{\rho}_0,\tilde{H}\hat{\rho}_0)\end{split}\label{a28}
\end{align}
implying
\begin{align}
\begin{split}
(\Delta\tilde{H})_0^2&=\frac{1}{\Vert\hat{\rho}_0\Vert^2}(\hat{\rho}_0,\tilde{H}^2\hat{\rho}_0)=\\
&=\frac{1}{\Vert\hat{\rho}_0\Vert^2}(\tilde{H}\hat{\rho}_0,\tilde{H}\hat{\rho}_0)=\frac{-Tr([\hat{H},\hat{\rho}_0]^2)}{Tr(\hat{\rho}_0^2)}.\end{split}\label{a29}
\end{align}
Eq.~\eqref{a23} yields then
\begin{equation}
\frac{Tr(\hat{\rho}_0\hat{\rho}_t)}{Tr(\hat{\rho}_0^2)}\geq\cos\naw{\sqrt{\frac{-Tr([\hat{H},\hat{\rho}_0]^2)}{Tr(\hat{\rho}_0^2)\hbar^2}}\, t}.\label{a30}
\end{equation}
By comparying \eqref{a30} with eqs.~\eqref{a16} and \eqref{a18} we conclude that it provides the quantum couterpart of \eqref{a16}.

Defining $\hat{\rho}^\alpha$, $\alpha\geq 0$, with the help of the spectral decomposition of $\hat{\rho}$ we find that it obeys the equation of motion \eqref{a19}. Therefore, assuming that $\hat{\rho}^\alpha$ is again a Hilbert-Schmidt operator we arrive at the quantum counterpart of eq.~\eqref{a17}.
Concluding, let us note that the bound \eqref{a30} is stronger that the ones obtained in  \cite{Shanahan} (cf.~the Appendix).

One can pose the question whether the bound \eqref{a30} is attainable. If $\hat{\rho}=\ket{\Psi}\bra{\Psi}$ is a pure state, eq.~\eqref{a30} implies
\begin{equation}
\modu{\av{\Psi(0)|\Psi(t)}}^2\geq\cos\naw{\frac{\sqrt{2}(\Delta E)_0t}{\hbar}}\label{a31}
\end{equation}  
while the original Mandelstam-Tamm bound yields
\begin{equation}
\modu{\av{\Psi(0)|\Psi(t)}}^2\geq\cos^2\naw{\frac{(\Delta E)_0t}{\hbar}}\label{a32}
\end{equation}
which is stronger \cite{Bengtsson}. So \eqref{a31} cannot be saturated.

Within our framework based on $\mathcal{H}_{HS}$, the normalized density matrix $\frac{1}{\Vert\hat{\rho}\Vert}\hat{\rho}$ is always viewed as a pure state. It is known that for pure states the Mandelstam-Tamm bound is attainable in the sense that, given two states, $\ket{\Psi}\equiv\ket{\Psi(0)}$, $\ket{\phi}\equiv\ket{\Psi(t)}$, one can find the Hamiltonian saturating the inequality \eqref{a32}. It should be selected in such a way as to generate the arc of the great circle connecting $\ket{\Psi(0)}$ and $\ket{\Psi(t)}$ on $S^2$ \cite{Bengtsson}. It reads
\begin{equation}
\tilde{H}=\omega(\ket{\Psi}\langle\tilde{\Psi}|+|\tilde{\Psi}\rangle\bra{\Psi})\label{a33}
\end{equation}
where
\begin{equation}
|\tilde{\Psi}\rangle=i\naw{\frac{\ket{\phi}-\ket{\Psi}\av{\Psi|\phi}}{\sqrt{1-\modu{\av{\phi|\Psi}}^2}}}\label{a34}
\end{equation}
i.e. $\langle\Psi|\tilde{\Psi}\rangle=0$, $\langle\tilde{\Psi}|\tilde{\Psi}\rangle=1$.

However, the trouble is that in our case the set of admissible Hamiltonians is restricted to those satisfying eq.~\eqref{a21}. We have checked explicitly that the Hamiltonians obeying eq.~\eqref{a21} cannot be chosen in the form described by eqs.~\eqref{a33} and \eqref{a34} (see Appendix B). Therefore, it is unlikely that the bound \eqref{a30} can be saturated for any state, pure or mixed. 

\section{Speed limit in terms of Wigner's functions}\label{IV}
As we have discussed in the Introduction one can expect that the smooth classical limit for the speed bound exists rather for strongly mixed states than pure ones. This is best seen if one uses the Wigner function formalism (the description of quantum speed limit in the framework of Wigner's function has been discussed by a number of authors cf., e.g., \cite{Defner2},\cite{Shanahan}).
Assume our space of states $\mathcal{H}$ describes the quantum system obtained by quantizing some classical Hamiltonian dynamics (for simplicity, we assume one degree of freedom). It is convenient to introduce a specific basis in the space $\mathcal{H}_{HS}$. Following Mukunda \cite{Mukunda} we define the operators $\omega(\hat{q})$ and $\omega(\hat{p})$, acting in $\mathcal{H}_{HS}$, by 
\begin{equation}
\omega(\hat{q})\hat{A}=\frac{1}{2}(\hat{q}\hat{A}+\hat{A}\hat{q})\label{a38}
\end{equation} 
\begin{equation}
\omega(\hat{p})\hat{A}=\frac{1}{2}(\hat{p}\hat{A}+\hat{A}\hat{p}).\label{a39}
\end{equation} 
They are selfadjoint with respect to the scalar product (\ref{a20}). Moreover, 
\begin{equation}
\com{\omega(\hat{q}),\omega(\hat{p})}=0.\label{a40}
\end{equation}
Therefore, their common eigenvectors $\hat{V}(q,p)$,
\begin{equation}
\omega(\hat{q})\hat{V}(q,p)=q\hat{V}(q,p)\label{a41}
\end{equation}
\begin{equation}
\omega(\hat{p})\hat{V}(q,p)=p\hat{V}(q,p)\label{a42}
\end{equation}
span the (generalized) basis in $\mathcal{H}_{HS}$. The solution to \eqref{a41}, \eqref{a42} reads
\begin{equation}
\hat{V}(q,p)=\sqrt{\frac{2}{\pi\hbar}}\,e^{\frac{2i}{\hbar}(p\hat{q}-q\hat{p})}P\label{a43}
\end{equation}
where $P$ is the parity operator,
\begin{equation}
P\ket{q}=\ket{-q}
\end{equation}
\begin{equation}
P\ket{p}=\ket{-p}.
\end{equation}
Moreover,
\begin{align}
\begin{split}
\naw{\hat{V}(q,p),\hat{V}(q',p')}&\equiv Tr\naw{\hat{V}^+(q,p)\hat{V}(q',p')}=\\
&=\delta(q-q')\delta(p-p').\end{split}
\end{align}
For any $\hat{A}\in\mathcal{H}_{HS}$ one has the eigenfunctions expansion
\begin{equation}
\hat{A}=\int \mathrm{d}q\,\mathrm{d}p\,\tilde{a}(q,p)\hat{V}(q,p). \label{a47}
\end{equation}
In particular,
\begin{equation}
Tr(\hat{A}^+\hat{B})=\int \mathrm{d}q\,\mathrm{d}p\,\overline{\tilde{a}(q,p)}\,\tilde{b}(q,p). \label{a48}
\end{equation}
and
\begin{equation}
\tilde{a}(q,p)=Tr(\hat{V}^+(q,p)\hat{A}).\label{a49}
\end{equation}
Computing the trace yields
\begin{equation}
\tilde{a}(q,p)=\frac{1}{\sqrt{2\pi\hbar}}\int\limits_{-\infty}^\infty \mathrm{d}q'\,\bra{q+\frac{q'}{2}}\hat{A}\ket{q-\frac{q'}{2}} e^{-\frac{ipq'}{\hbar}}.\label{a50}
\end{equation}
Let us comment on the formulae \eqref{a38}$\div$\eqref{a50}. Our main idea is to obtain the relevant bound(s) on evolution speed by passing from the initial space of states $\mathcal{H}$ by the Hilbert space $\mathcal{H}_{HS}$ of Hilbert-Schmidt operators acting in $\mathcal{H}$; all, pure and mixed, states in $\mathcal{H}$ become pure states in $\mathcal{H}_{HS}$ which allows to apply immediately the original Mandelstam-Tamm arguments. Eqs.~\eqref{a38}$\div$\eqref{a50} show clearly that the "coordinate" representation in $\mathcal{H}_{HS}$ can be obtained by viewing the original phase space as new configuration space with $\hat{V}(q,p)$ spanning the coordinate basis (cf.~eqs.~\eqref{a41}, \eqref{a42} and \eqref{a47}). In particular, the Wigner function defined by
\begin{align}\begin{split}
W&(q,p)\equiv \frac{1}{\sqrt{2\pi\hbar}}\,\tilde{\rho}(q,p)=\\
&=\frac{1}{2\pi\hbar}\int\limits_{-\infty}^\infty \mathrm{d}q'\,\bra{q+\frac{q'}{2}}\hat{\rho}\ket{q-\frac{q'}{2}} e^{-\frac{ipq'}{\hbar}}.\end{split}\label{a51}
\end{align}
 becomes the coordinate wave function of the (now pure) state $\hat{\rho}$.\\
Eqs.~\eqref{a48} and \eqref{a51} imply
\begin{equation}
Tr(\hat{\rho}_0\hat{\rho}_t)=2\pi\hbar\int\mathrm{d}q\,\mathrm{d}p\,W(q,p,0)\,W(q,p,t).\label{a52}
\end{equation}
One readily concludes from eq.~\eqref{a50} that 
\begin{equation}
a_W(q,p)=\sqrt{2\pi\hbar}\,\tilde{a}(q,p)\label{a53}
\end{equation}
is the Weyl symbol of $\hat{A}$. In particular, the Wigner function is the Weyl symbol of density matrix $\hat{\rho}$ divided by $2\pi\hbar$. Therefore, puting $\hat{B}=\hat{\rho}$ in eq.~\eqref{a48} we find
\begin{equation}
\langle\hat{A}\rangle_\rho\equiv Tr(\hat{A}\hat{\rho})=\int\mathrm{d}q\,\mathrm{d}p\, W(q,p)\,a_W(q,p).\label{a54}
\end{equation}
The normalization condition reads
\begin{equation}
1=Tr\hat{\rho}=\int \mathrm{d}q\,\mathrm{d}p\, W(q,p)\label{c52}
\end{equation}
while 
\begin{equation}
1\geq Tr\hat{\rho}^2=2\pi\hbar\int\mathrm{d}q\,\mathrm{d}p\, W^2(q,p).\label{a56}
\end{equation}
Taking into account that in the limit $\hbar\rightarrow 0$ the Weyl symbol becomes the corresponding classical dynamical variable we conclude from eq.~\eqref{a54} that the Wigner function should be in this limit identified with classical probability density on phase space

\begin{equation}
\rho(q,p)=\lim_{\hbar\rightarrow 0}W(q,p).\label{a57}
\end{equation}

The normalization condition on classical probability distribution reads 
\begin{equation}
\int \text{d}q\text{d}p\,\rho(p,q)=1\label{54a}
\end{equation}
and follows immediately from eqs.~\eqref{c52} and \eqref{a57}. However, the speed limit for evolution in classical phase space \cite{Okuyama} (cf. Sec.~II) has been derived under the assumption that $\rho(q,p)$ belongs to the Hilbert space of functions square integrable over the phase space, i.e.
\begin{equation}
\int \text{d}q\text{d}p\,\rho^2(p,q)<\infty\label{54b}
\end{equation}
and \eqref{54a} does not imply \eqref{54b}. Therefore, one can expect the existence of classical counterpart of quantum speed limit only for quantum states yielding square integrable classical distributions. It follows then from eqs.~\eqref{a56}, \eqref{a57} and \eqref{54b} that 
\begin{equation}
\lim_{\hbar\rightarrow 0}Tr\hat{\rho}^2=0.\label{a58}
\end{equation}
Now, the pure states are uniquely characterized by the condition $Tr\hat{\rho}^2=1$. Eq.~\eqref{a58} implies that, with $\hbar\rightarrow 0$, $\hat{\rho}$ becomes a mixture of growing number of pure states entering with nonnegligible weights. To see this let $\poisson{\ket{n}}_{n=1}^\infty$ be the basis diagonalizing $\hat{\rho}$:
\begin{equation}
\hat{\rho}=\sum_{n=1}^{\infty}\ket{n}p_n\bra{n},\quad p_n\geq 0,\quad\sum_{n=1}^\infty p_n=1.
\end{equation}
Then 
\begin{equation}
Tr\hat{\rho}^2=\sum_{n=1}^\infty p_n^2.\label{a60a}
\end{equation}
Let us select some natural $N$ and let 
\begin{equation}
P_N\equiv \sum_{n=N+1}^\infty p_n;
\end{equation}
obviously $P_N\rightarrow 0$ as $N\rightarrow\infty$. Now, 
\begin{equation}
Tr\hat{\rho}^2\geq \sum_{n=1}^\infty p_n^2
\end{equation}
and the right-hand side can be minimalised under the constraint
$\sum\limits_{n=1}^N p_n=1-P_N$ to yield
\begin{equation}
Tr\hat{\rho}^2\geq\frac{(1-P_N)^2}{N}.\label{c60}
\end{equation}
For pure states, $p_n=1$ for some $n=n_0$ and $p_n=0$ otherwise. Therefore, the number $N$ of states which contribute to the mixture can be taken as a measure of the mixing degree. Eqs.~\eqref{a56} and \eqref{c60} imply that $N\rightarrow \infty$ as $\hbar\rightarrow 0$ (roughly, $N\sim\frac{1}{\hbar}$), i.e. more and more states with comparable weights should be included as $\hbar\rightarrow 0$. Note that for the above uniform distribution, $p_n\sim\frac{1}{N}$, $P_N\sim 0$, von Neumann entropy $S=-\sum\limits_{n=1}^N p_n\ln p_n$ attains the maximal value $S=\ln N\sim -\ln\hbar$.
We see that one cannot obtain square integrable classical probability distribution in the $\hbar\rightarrow 0$ limit for pure states; this continues to be the case for the mixtures of finite numbers of pure states. The square integrable classical probability distribution can only emerge provided the number of states contributing to the spectral decomposition of the density matrix grows indefinitely as $\hbar\rightarrow 0$.

A nice illustration is provided by the canonical distribution for harmonic oscillator. Then the $\ket{n}$ states are the eigenstates of the harmonic oscillator Hamiltonian while
\begin{equation}
p_n=\naw{1-e^{-\frac{\hbar\omega}{kT}}}e^{-\frac{n\hbar\omega}{kT}}\label{c63}
\end{equation} 
yielding
\begin{equation}
Tr\hat{\rho}^2=\frac{\naw{1-e^{-\frac{\hbar\omega}{kT}}}^2}{1-e^{-\frac{2\hbar\omega}{kT}}}.\label{c64}
\end{equation}
Now, in the "strongly quantum" (large $\hbar$)/low temperature (small $T$) regime one finds
\begin{equation}
Tr\hat{\rho}^2\simeq 1-2e^{-\frac{\hbar\omega}{kT}}
\end{equation}
 i.e. $Tr\hat{\rho}^2$ differs from unity by exponentially small term; the state is almost pure one. The physical reason for that is clear: the energy gap between the ground state and the first excited one is large as compared to that of thermal excitations, $\hbar\omega \gg kT$, so the probability that the oscillator stays in the ground state is exponentially close to unity.
 
 On the other hand, in the semiclassical (small $\hbar$)/high temperature (large $T$) regime more and more excited states contribute to $\hat{\rho}$ with comparable weights; the energy levels differences become small as compared to the scale of thermal fluctuations. Eq.~\eqref{c64} yields now
 \begin{equation}
 Tr\hat{\rho}^2\simeq\frac{\hbar\omega}{2kT}\underset{\hbar\rightarrow
  0}\longrightarrow 0.
\end{equation}  
It is easy to compute the relevant Wigner function. Obviously
\begin{equation}
W(q,p)=\sum_{n=0}^\infty p_nW_n(q,p)
\end{equation} 
with $W_n(q,p)$ being the Wigner function for $n$-th energy eigenstate. Using \eqref{c63} and the explicit form of $W_n(q,p)$ (see, for example, \cite{Shanahan}) one finds
\begin{equation}
W(q,p)=\frac{1}{\pi\hbar}\tanh\naw{\frac{\hbar\omega}{2kT}}e^{-\frac{\tanh\naw{\frac{\hbar\omega}{2kT}}}{\hbar\omega}(p^2+\omega^2q^2)}
\end{equation}
and
\begin{equation}
\lim_{\hbar\rightarrow 0}W(q,p)=\frac{\omega}{2\pi kT}e^{-\frac{p^2+\omega^2q^2}{2kT}}
\end{equation}
in accordance with eq.~\eqref{a57}. The limiting distribution (classical canonical distribution) is square integrable and is obtained by including more and more eigenstates as $\hbar\rightarrow 0$ ($n\hbar\sim O(1)$) in accordance with the correspondence principle.

The conclusion concerning the growing degree of mixing supports our general discussion in Sec. \ref{I}.

In order to rewrite the right hand side of eq.~\eqref{a30} in terms of Wigner's function let us remaind that
\begin{equation}
\naw{\frac{1}{i\hbar}[\hat{A},\hat{B}]}_W=\poisson{\poisson{a_W,b_W}}
\end{equation}
where on the left hand side $(\ldots)_W$ denotes the Weyl symbol of the expression in bracket while $\{\{\;,\;\}\}$ on the right hand side denotes the Moyal bracket \cite{Moyal}, \cite{Zachos}. Using eqs.~\eqref{a48}, \eqref{a51} and \eqref{a53} we find
\begin{equation}
-\frac{1}{\hbar^2}Tr\naw{[\hat{H},\hat{\rho}_0]^2}=2\pi\hbar\int\mathrm{d}q\,\mathrm{d}p\poisson{\poisson{H_W,W_0}}^2 
\end{equation}
where $W_0\equiv W(q,p,0)$ and $H_W$ denotes the Weyl symbol of the Hamiltonian $\hat{H}$. Finally, taking into account eq.~\eqref{a52} one can rewrite \eqref{a30} as
\begin{align}
\begin{split}
&\frac{\int\mathrm{d}q\,\mathrm{d}p\,W(q,p,0)\,W(q,p,t)}{\int\mathrm{d}q\,\mathrm{d}p\,W(q,p,0)^2}\geq\\
&\geq\cos\naw{\sqrt{\frac{\int\mathrm{d}q\,\mathrm{d}p\,\poisson{\poisson{H_W,W_0}}^2}{\int\mathrm{d}q\,\mathrm{d}p\,W_0^2}}\,\cdot t}.\end{split}\label{a59}
\end{align}
Note that
\begin{equation}
\poisson{\poisson{.\,,.}}=\poisson{.\,,.}+O(\hbar^2)\label{a60}
\end{equation}
where $\poisson{.\,,.}$ denotes the classical Poisson bracket. By comparying eqs.~\eqref{a59} and \eqref{a60} with \eqref{a16} and \eqref{a18} we conclude that eq.~\eqref{a16} coincides with $\hbar\rightarrow 0$ limit of eq.~\eqref{a59}.

\section{Summary}\label{V}
We have derived the quantum bound on relative purity. It is given by eqs.~\eqref{a30} and is tighter than the one encountered in this context in the literature (see Appendix). Its classical limit coincides with the bound derived by Okuyama and Ohzeki. The existence of classical limit is related to the degree of mixing of the state under consideration: it becomes more and more mixed as $\hbar\rightarrow 0$.

Actually, as in the classical case, we have the whole family of bounds. One can replace $\hat{\rho}$ by $\hat{\rho}^\alpha$, $\alpha\in\mathbbm{R}_+$, provided the latter is well defined (unnormalized) density operator.

The final remark we would like to make concerns the systems described by finitedimensional Hilbert spaces ("spin" systems). In order to compare the classical and quantum speed limits one has to define carefully the $\hbar\rightarrow 0$ limit for such systems. Typically, this involves considering larger and larger dimensionality $N$ of the space of states with $N\hbar$ fixed. Therefore, again $\text{Tr}(\rho^2)$ can vanish in the limit $\hbar\rightarrow 0$. However, we postpone a more detailed study of this issue for further considerations.
\subsection*{Acknowledgement}
We are grateful to Profs.~Krzysztof Andrzejewski, Cezary Gonera and Pawe\l{} Ma\'slanka for fruitful discussion.

\bibliographystyle{plain}

\onecolumn\newpage
\appendix
\section{Appendix}
In order to compare our bound on the speed of quantum evolution let us remind the derivation of the original MT bound (cf.~nice discussion in Ref.~\cite{Wootters}). Let
\begin{equation}
\rho(t)\equiv\ket{\Psi(t)}\bra{\Psi(t)}
\end{equation} 
be the density matrix of the pure state $\ket{\Psi(t)}$. We are interested in the behaviour of the quantity (relative purity/fidelity) defined by 
\begin{equation}
F(t)\equiv\modu{\av{\Psi(0)\vert\Psi(t)}}^2=\text{Tr}\naw{\rho(0)\rho(t)}\equiv\av{\rho(t)}_0.
\end{equation}
It obeys 
\begin{equation}
\dot{F}(t)=\frac{1}{i\hbar}\av{\com{H,\rho(t)}}_0.\label{a72}
\end{equation}
The uncertainty principle yields
\begin{equation}
\modu{\av{\com{H,\rho(t)}}}_0\leq(\Delta H)_0\cdot(\Delta\rho(t))_0
\end{equation}
where $(\Delta H)_0^2\equiv\av{H^2}_0-\av{H}^2_0$ etc. In particular,
\begin{equation}
\naw{\Delta\rho(t)}_0^2=\av{\rho^2(t)}_0-\av{\rho(t)}_0^2=\av{\rho(t)}_0-\av{\rho(t)}_0^2\equiv F(t)-F^2(t).\label{a74}
\end{equation}
Eqs.~\eqref{a72}$\div$\eqref{a74} give
\begin{equation}
\vert\dot{F}(t)\vert\leq\frac{2}{\hbar}(\Delta E)_0\sqrt{F(t)-F^2(t)}
\end{equation}
Integrating the last inequality one finds
\begin{equation}
\modu{\av{\Psi(t)\vert\Psi(0)}}^2\geq\cos^2\naw{\frac{(\Delta E)_0}{\hbar}\cdot t}.\label{a76}
\end{equation}
On the other hand one can use the Cauchy-Schwarz inequality in a more straightforward way. Rewriting the right hand side of \eqref{a72} as follows
\begin{equation}
\frac{1}{i\hbar}\av{\com{H,\rho(t)}}_0=\frac{1}{i\hbar}\text{Tr}\naw{\rho(0)\com{H,\rho(t)}}=-\frac{1}{i\hbar}\text{Tr}\naw{\rho(t)\com{H,\rho(0)}}
\end{equation}
and applying Cauchy-Schwarz inequality  $\modu{\text{Tr}(A^+B)}\leq\sqrt{\text{Tr}(A^+A)}\sqrt{\text{Tr}(B^+B)}$ one finds immediately
\begin{equation}
\vert\dot{F}(t)\vert\leq\frac{1}{\hbar}\sqrt{-\text{Tr}\naw{\com{\rho(0),H}^2}}=\frac{\sqrt{2}}{\hbar}(\Delta E)_0
\end{equation}
or, in the integrated form 
\begin{equation}
\modu{\av{\Psi(t)\vert\Psi(0)}}^2\geq 1-\frac{\sqrt{2}}{\hbar}(\Delta E)_0\, t.\label{a79}
\end{equation}
This is exactly the estimate (12) from Ref.~\cite{Shanahan}.

\begin{figure}
\centering\includegraphics[scale=0.6]{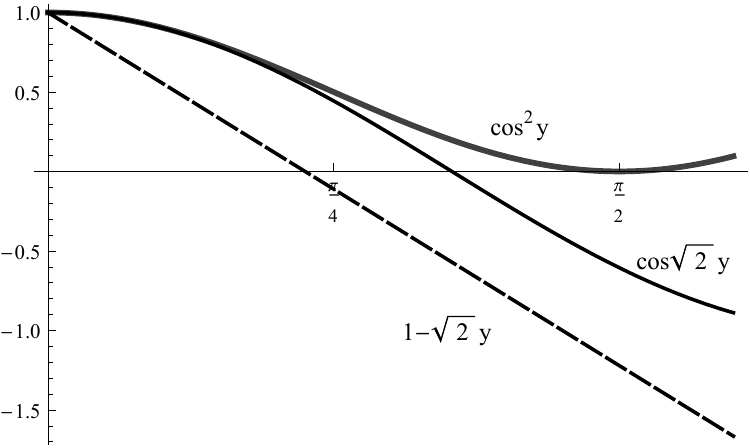}
\caption{Comparison of the bounds \eqref{a31}, \eqref{a76} and \eqref{a79}}\label{f1}
\end{figure}

Let us compare the bounds \eqref{a76} and \eqref{a79} with the one derived in the main text, eq.~\eqref{a31}. The right hand sides of eqs.~\eqref{a31}, \eqref{a76} and \eqref{a79} are depicted on Figure \ref{f1} below (with $y\equiv\frac{(\Delta E)_0}{\hbar}\,t$).
It is clearly seen that our estimate is considerably better than that presented in Ref.~\cite{Shanahan}. It is also true that the original MT bound is the best among the above ones. However, both our bound and the one derived by Shanahan et al.~\cite{Shanahan} generalize immediately to the mixed states. As we have shown mixed states must be taken into account when considering the $\hbar\rightarrow 0$ limit.

Also on the purely classical level the estimates considered here are stronger than those considered in Ref.~\cite{Shanahan}. Eq.~\eqref{a17} (derived originally by Okuyama and Ohzeki \cite{Okuyama}) yields for $\alpha=\frac{1}{2}$ the stronger inequality than eq.~(24) in Ref.~\cite{Shanahan}. 

\section{Appendix}
We show here that the construction leading to the Hamiltonian saturating the inequality \eqref{a32} cannot be performed in our case due to the fact that the set of admissible Hamiltonians is restricted to those satisfying eq. \eqref{a21}. To this end let us note that eq. \eqref{a33} implies
\begin{equation}
\tilde{H}\ket{\Psi}=\omega\ket{\tilde{\Psi}},\quad \tilde{H}\ket{\tilde{\Psi}}=\omega\ket{\Psi}.\label{c84}
\end{equation}
Now, in the context of $\mathcal{H}_{HS}$,
\begin{equation}
\ket{\Psi}=\frac{1}{\parallel\hat{\rho}_0\parallel}\hat{\rho}_0,\quad \ket{\phi}=\frac{1}{\parallel\hat{\rho}_t\parallel}\hat{\rho}_t.
\end{equation}
Moreover, the Hamiltonian $\tilde{H}$ acts according to the eq.~\eqref{a21} with $\hat{H}$ being an operator acting in $\mathcal{H}$. Therefore, eqs.~\eqref{c84} take the form
\begin{equation}
\begin{split}
&\com{\hat{H},\hat{\rho}_0}=i\omega\naw{\frac{\parallel\hat{\rho}_0\parallel^2\hat{\rho}_t-Tr(\hat{\rho}_0\hat{\rho}_t)\hat{\rho}_0}{\sqrt{\parallel\hat{\rho}_0\parallel^2\parallel\hat{\rho}_t\parallel^2-Tr^2(\hat{\rho}_0\hat{\rho}_t)}}}\\
& \com{\hat{H},\frac{\parallel\hat{\rho}_0\parallel^2\hat{\rho}_t-Tr(\hat{\rho}_0\hat{\rho}_t)\hat{\rho}_0}{\sqrt{\parallel\hat{\rho}_0\parallel^2\parallel\hat{\rho}_t\parallel^2-Tr^2(\hat{\rho}_0\hat{\rho}_t)}}}=-i\omega\hat{\rho}_0
\end{split}\label{c86}
\end{equation}
Taking the traces of both sides of eqs.~\eqref{c86} and using $Tr\hat{\rho}_0=1=Tr\hat{\rho}_t$ one finds that they are contradictory except the case $\hat{\rho}_0=\hat{\rho}_t$ when they become indefinite.

\end{document}